# Mapping the nonlinear optical susceptibility by noncollinear second harmonic generation

M.C.Larciprete,[1,*] F.A.Bovino,[2] M.Giardina,[2] A. Belardini,[1] M. Centini,[1] C. Sibilia,[1] M. Bertolotti,[1] A.Passaseo,[3] V.Tasco,[3]

[1] *Dipartimento di Energetica – Università di Roma "La Sapienza"*
*Via A. Scarpa 16, 00161 Roma, ITALY.*

[2] *Quantum Optics Lab. , Elsag-Datamat Via Puccini 2 Genova, ITALY.*

[3] *CNR-NNL-INFM Unità di Lecce - Via per Arnesano Lecce, ITALY*

[*]*Corresponding author: mariacristina.larciprete@uniroma1.it*

We present a method, based on noncollinear second harmonic generation, to evaluate the non-zero elements of the nonlinear optical susceptibility. At a fixed incidence angle, the generated signal is investigated by varying the polarization state of both fundamental beams. The resulting polarization charts allows to verify if Kleinman's symmetry rules can be applied to a given material or to retrieve the absolute value of the nonlinear optical tensor terms, from a reference measurement. Experimental measurements obtained from Gallium Nitride layers are reported. The proposed method doesn't require an angular scan thus is useful when the generated signal is strongly affected by sample rotation. 2008 Optical Society of America

Noncollinear second harmonic generation (SHG) was firstly developed by Muenchausen [1] and Provencher [2]. More recently, Faccio [3] pointed out that this process presents the advantage of a reduced coherence length while both Figliozzi [4] and Cattaneo [5] have shown that the technique allows the bulk and surface responses to be addressed. In addition, Cattaneo has also demonstrated that the technique is very useful in surface and thin-film characterization [6-7]. Given the two pump beams tuned at $\omega_1=\omega_2=\omega$, having different incidence angles, $\alpha_1$ an $\alpha_2$, and polarization angles, $\phi_1$ an $\phi_2$, defined with respect to the y-z plane, several experimental parameters can be combined to excite a noncollinear nonlinear optical polarization $P(2\omega)^{(2)} = \chi^{(2)} : E_1(\omega)E_2(\omega)$.

We employed noncollinear SHG to map the generated signal as a function of both pump beams polarization state, from Gallium Nitride slab, 302 nm thick, grown by metal-organic chemical vapour deposition (MOCVD) onto (0001) $c$-plane $Al_2O_3$ substrates [8]. GaN presents a wurtzite crystal structure without centre of inversion, thus leading to efficient second order nonlinear effects [9-10].

The obtained results show the peculiarity of the nonlinear optical response associated with noncollinear excitation, and are fully explained using the expression for the effective second order optical nonlinearity in noncollinear scheme. We show that the polarization charts, at a fixed rotation angle of the sample, offer the possibility to evaluate the ratio between the different non-zero elements of the nonlinear optical tensor, thus verifying if Kleinman's symmetry rules can be applied to a given material. Moreover, if the measurements are performed with reference to a well-characterized sample, i.e. quartz or KDP, this method allows the evaluation of the absolute values of the non-zero terms of the nonlinear optical tensor, without requiring sample rotation. As a consequence, the proposed technique results particularly appropriate for those conditions where the generated signal can be strongly affected by sample rotation angle. For samples which are some coherence lengths thick, as the optical path length is changed by rotation, the SH signal strongly oscillates with increasing incidence angle [11] according to Maker-fringes pattern, thus a high angle resolution is required. When using short laser pulses, whose bandwidth is comparable or lower than sample thickness, as the incidence angle is modified the nonlinear interaction may involve different part of the sample and, eventually, surface contributions. For nano-patterned samples, finally, a rotation would imply differences into sample surface interested by the pump spot size. With respect to the mentioned examples, the method of polarization scan simplifies the characterization of the nonlinear optical tensor elements without varying the experimental conditions.

Leaving aside third and higher harmonics, the resulting nonlinear polarization is composed by three waves tuned at $2\omega_1$, $2\omega_2$, and $\omega_1+\omega_2$. While the first two waves are nearly collinear with the incident beams, propagating at the internal angles $\alpha'_1$ and $\alpha'_2$, the third one, given the conservation of wavevector tangential components, $k_{\omega 1} sin(\alpha'_1) + k_{\omega 2} sin(\alpha'_2) = k_{\omega 1+\omega 2} sin(\alpha'_3)$, is emitted approximately along the bisector of the aperture angle between the two pump beams, $\alpha'_3$.

Wurtzite crystal structure, presents the noncentrosymmetric point group symmetry 6mm with a hexagonal primary cell. The second order susceptibility tensor [12-13] presents only two independent nonvanishing coefficients in wavelength regimes where it is possible to take advantage of Kleinmann symmetry rules, i.e. $d_{15}=d_{24}=d_{31}=d_{32}$ and $d_{33}$; furthermore, for an ideal wurtzite structure the nonzero elements are related to each other via the $d_{33} = -2 \cdot d_{31}$ [14].

The full expression of the generated SH power is a function of sample rotation angle $\alpha$, as well as propagation angle and polarization state of both fundamental and generated beams, trhough the Fresnel coefficients and includes the effective susceptibility, $d_{eff}(\alpha)$ [15]. In the case of GaN, given the two incident fields as $\vec{E}_i = (sin(\phi_i) \; -cos(\phi_i)cos(\alpha_i) \; -cos(\phi_i)sin(\alpha_i))$, the expressions for $d_{eff}(\alpha)$ as a function of $\alpha'_1$ and $\alpha'_2$, the noncollinear SH beam propagation angle $\alpha'_3$, and polarization state, $\phi_1$ and $\phi_2$ is:

$$d^S_{eff} = -d_{15}[sin(\phi_1)cos(\phi_2)sin(\alpha'_2) + cos(\phi_1)sin(\phi_2)sin(\alpha'_1)] \qquad (1)$$

$$d^P_{eff} = -d_{24}cos(\alpha'_3)[cos(\alpha'_1)sin(\alpha'_2) + sin(\alpha'_1)cos(\alpha'_2)]cos(\phi_1)cos(\phi_2) -$$

$$- sin(\alpha'_3)\{d_{31}sin(\phi_1)sin(\phi_2) + cos(\phi_1)cos(\phi_2)[d_{32}cos(\alpha'_1)cos(\alpha'_2) + d_{33}sin(\alpha'_1)sin(\alpha'_2)]\}$$

where the apex *S* or *P* stands for the polarization state of the generated beam.

The output of a mode-locked Ti:Sapphire laser system tuned at λ=830 nm (76 MHz repetition rate, 130 fs pulse width) was split into two beams of about the same intensity. The polarization of both beams can be varied with two identical half wave plates. The sample was placed onto a rotation stage allowing the variation of the sample rotation angle, α, with a resolution of 0.05 degrees. After passing through two lenses, 150 mm focal length, the pump beams were sent to intersect in the focus region with the angles β=9° and γ= -9°, measured with respect to α = 0°. Thus for a given α ≠ 0°, the corresponding incidence angles of the two pump beams result to be $\alpha_1=\alpha-\beta$ and $\alpha_2=\alpha-\gamma$ (see Figure 1). The noncollinear SH beam was focused with an objective on to a monomodal optical fiber coupled with a photon counting detector. A set of short wave pass filters was used to suppress any residual light at $\omega_1$ and $\omega_2$, while an analyzer allowed to select the desired SH polarization state. Finally, the temporal overlap of the incident pulses was controlled with an external delay line.

In the experiments, the two half-wave plates were rotated, in the range -180° ÷ +180° degrees for the first pump beam ($\phi_1$) and 0° ÷ 180° degrees for the second pump beam ($\phi_2$), with a resolution of 4° degrees. The resulting experimental curves systematically describe all possible combinations for the polarization state of the two pump beams. The results obtained at α= 35° for the two different polarization state of the analyzer, namely $\hat{p}$, $\phi$=0°, and $\hat{s}$, $\phi$=90° are shown in Figures 2 and 3.

The experimental surface plots of the noncollinear SH signal generated in $\hat{p}$ polarization (Figure 2a), show that the absolute maxima are achievable when both pumps are $\hat{p}$-polarized, i.e. when $\phi_1$ and $\phi_2$ are both 0° or 180°, while relative maxima occur when both pumps are $\hat{s}$-polarized, i.e. when polarization angles of both pumps are set to ± 90°. Conversely, when the two pump beams have crossed polarization, i.e when $\phi_1$=0° and $\phi_2$=90° and viceversa, the nonlinear optical tensor do not allow SH signal which is $\hat{p}$-polarized. Measurements performed at different sample rotation angles, displaied a similar symmetry, while amplitude is decreasing with decreasing rotation angle. A fairly different behavior is observable, when the analyzer is set to $\hat{s}$-polarization, i.e. $\phi$=90°. In this case the maxima generally occur when the two pump beams have crossed polarization. Experimental measurements (Figure 3a), at the same α=35°, show that the absolute maxima take place when the first pump is $\hat{s}$-polarized and the second pump is $\hat{p}$-polarized, i.e. $\phi_1$= ±90° and $\phi_2$ is equal to either 0° or 180°. Relative maxima occur in the reverse situation, when the first pump is $\hat{p}$-polarized, $\phi_1$= 0° or ±180°, and the second pump $\hat{s}$-polarized, $\phi_2$= 90°. When the two pumps are equally polarized, either $\hat{s}$ or $\hat{p}$, there is no SH signal $\hat{s}$-polarized. Due to the experimental geometry, this condition is not symmetrical for positive and negative rotation angles, thus the resulting surface plots obtained at different rotation angles present some variations.

The calculated polarization charts, reported in Figure 2b and 3b, were retrieved from equation 1 by assuming for the nonlinear optical tensor elements the Kleinmann symmetry rules. The perfect matching between the experimental and the theoretical charts verify the rightness of the symmetry assumption. Assuming a different relationship between the coefficients $d_{15}$, $d_{31}$ and $d_{33}$ would in fact lead to evident modifies in the polarization charts.

In conclusion, we studied noncollinear SHG at a fixed incidence angle by investigating the generated signal arising from all the possible combinations of the polarization states of the two pump beams. Our experimental and theoretical results show that the obtainable polarization charts, offer all the information to evaluate the ratio between the different non-zero elements of the nonlinear optical tensor. The method we have described is an alternative method with respect to Maker fringes technique, applied to the noncollinear case and represents a useful tool to characterize the non-zero terms of the nonlinear optical tensor without varying one of the relevant experimental conditions as the incidence angle.

**Figure Captions**

**Fig. 1.** Details of the experimental geometry used for noncollinear second harmonic generation. For a given rotation angle α, the corresponding incidence angles of the two pump beams result to be $\alpha_1=\alpha-\beta$ and $\alpha_2=\alpha-\gamma$. Black arrows indicate the crystalline axes.

**Fig. 2.** Second harmonic signal as a function of the polarization state of the first pump beam ($\phi_1$) and the second pump beam ($\phi_2$), (a) experimentally measured and (b) theoretically calculated. Sample rotation angle was fixed to α=35°. The polarization state of the analyzer is set to $\hat{p}$, i.e. $\phi=0°$. The inset is a schematic representation of the polarization of pump beams at the absolute maxima, i.e. when both pump beams are p-polarized $\hat{p}$.

**Fig. 3.** Second harmonic signal as a function of the polarization state of the first pump beam ($\phi_1$) and the second pump beam ($\phi_2$), (a) experimentally measured and (b) theoretically calculated. Sample rotation angle was fixed to α=35°. Polarization state of the analyzer is set to $\hat{s}$, i.e. $\phi=90°$. The inset show the polarization of pump beams at the absolute maxima, i.e. when the first pump is $\hat{s}$-polarized and the second pump is $\hat{p}$-polarized.

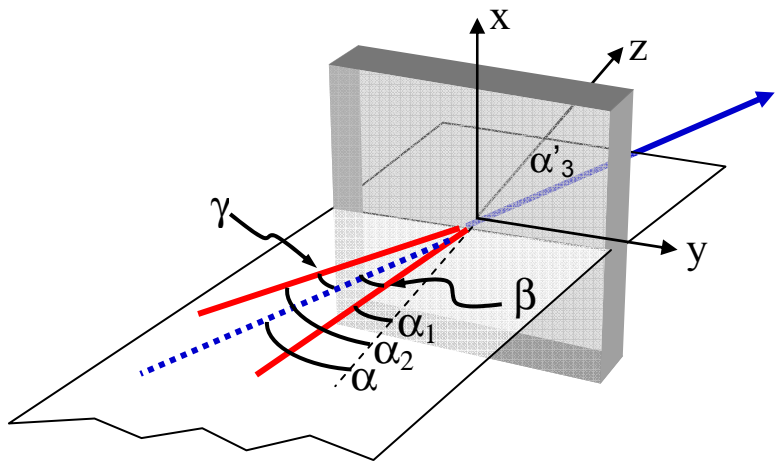

**Figure 1**

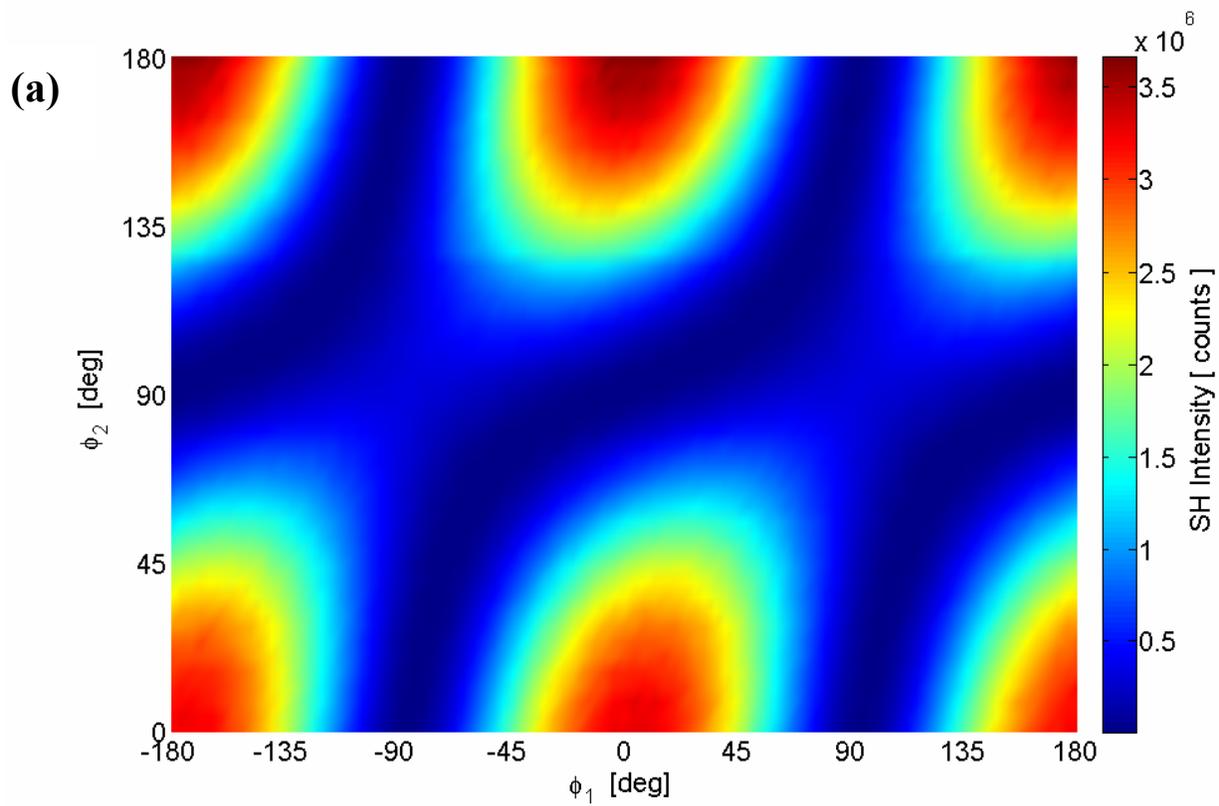
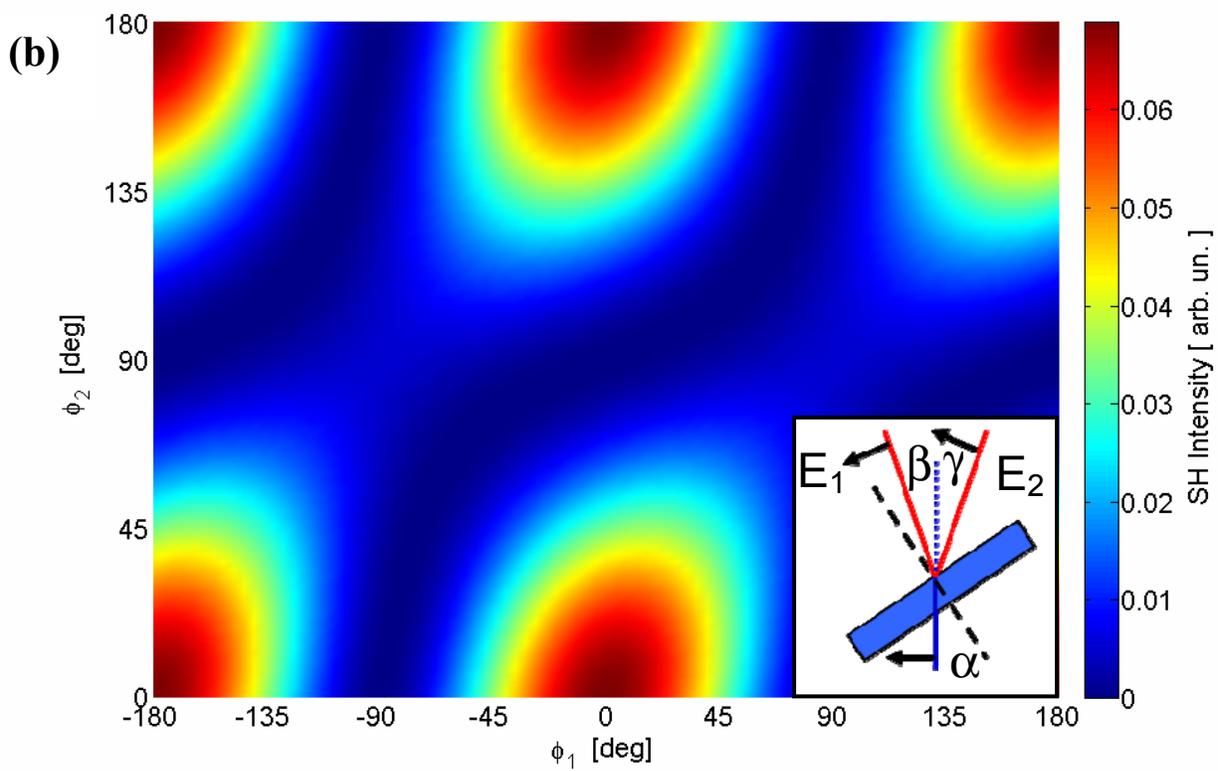

**Figure 2**

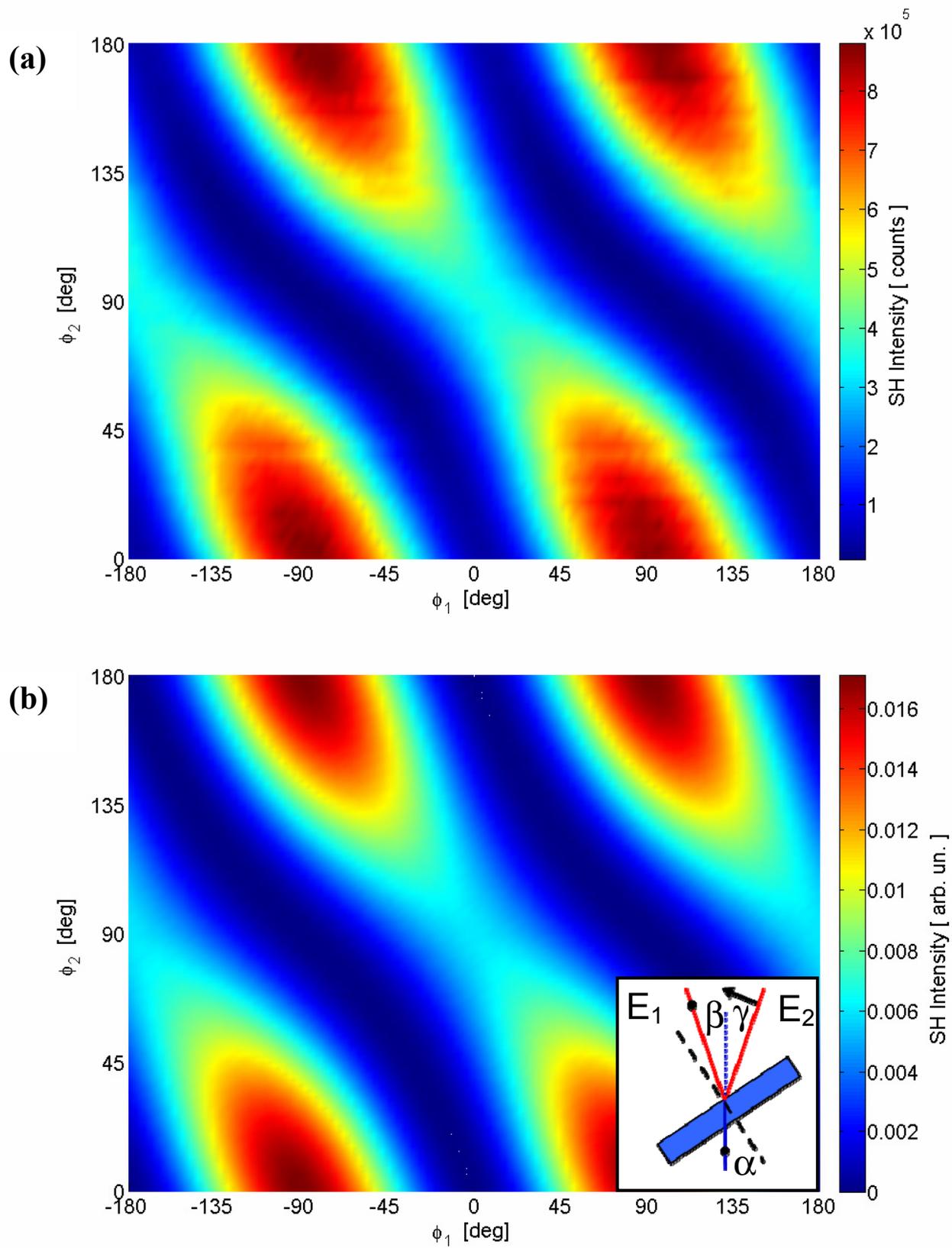

**Figure 3**